\documentclass[pre%
,twocolumn%
,superscriptaddress%
,floats
,aps%
,amssymb%
]{revtex4}
\usepackage{amsmath}%

\begin{document}
\title{Reduction of Spin Glasses applied to the Migdal-Kadanoff Hierarchical Lattice} 
\date{\today}
\author{Stefan Boettcher}  
\email{www.physics.emory.edu/faculty/boettcher}
\affiliation{Physics Department, Emory University, Atlanta, Georgia
30322, USA}  

\begin{abstract} 
A reduction procedure to obtain ground states of spin glasses on
sparse graphs is developed and tested on the hierarchical lattice
associated with the Migdal-Kadanoff approximation for low-dimensional
lattices. While more generally applicable, these rules here lead to a
complete reduction of the lattice. The stiffness exponent governing
the scaling of the defect energy $\Delta E$ with system size $L$,
$\sigma(\Delta E)\sim L^y$, is obtained as $y_3=0.25546(3)$ by
reducing the equivalent of lattices up to $L=2^{100}$ in $d=3$, and as
$y_4=0.76382(4)$ for up to $L=2^{35}$ in $d=4$. The reduction rules
allow the exact determination of the ground state energy, entropy, and
also provide an approximation to the overlap distribution. With these
methods, some well-know and some new features of diluted hierarchical
lattices are calculated.
\hfil\break  PACS number(s): 
05.50.+q
, 75.10.Nr
, 02.60.Pn
.
\end{abstract} 
\maketitle

\section{Introduction}
\label{intro}
\begin{figure}
\vskip 1.1in \includegraphics{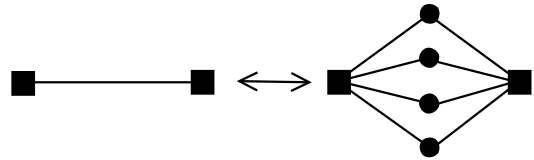}
\caption{Recursive generation of the hierarchical lattice,
proceeding from left to right. The reduction rules proceed
from right to left, replacing the sub-graph with a reduced bond. }
\label{hierlat}
\end{figure}

We propose a set of reduction rules applicable to spin glasses at
$T=0$ on any sparse graph and arbitrary bond distribution. These
reductions strip graphs of all variables that are connected to at most
two neighbors while accounting {\it exactly} for the ground state
energy, entropy, and approximately for the overlap distribution of the
system. In this paper, we introduce these reduction rules, and test
them on the hierarchical lattice (see Fig.~\ref{hierlat}) obtained
from the Migdal-Kadanoff bond-moving scheme \cite{MK}.  The recursive
nature of the hierarchical lattice permits us to quickly reduce the
equivalent of $10^9$ graphs with lengths corresponding to $L=2^{100}$
in $d=3$ and $L=2^{35}$ in $d=4$ dimensions, limited only by
accumulating rounding errors. Throughout this paper we focus
exclusively on $d=3$ and~4, and use only a discrete bond distribution,
\begin{eqnarray}
P(J)=p\,\delta(J^2-1)+(1-p)\,\delta(J).
\label{bondeq}
\end{eqnarray}
But our procedure is equally applicable to any continuous bond
distribution, such as a Gaussian.

The reduction produces high-accuracy results for the scaling of the
defect energy width $\sigma(\Delta E)$ with $L$. We also study the
diluted lattice, where $p$ shall refer to the fraction of occupied
bonds, and show that this scaling emerges for all $p>p^*$. The
critical point $p^*=0.31032$~\cite{BF} is particular to the bond
distribution in Eq.~(\ref{bondeq}) and located just above the
percolation point $p_c=0.281837$ of the lattice. Since $p_c$ is a
purely topological property of the lattice itself, having $p>p_c$ is
merely a necessary condition for long-range correlated behavior in a
spin glass. Cooperative effects from the bond disorder suppress
correlations even for $p_c<p<p^*$, as was already discussed in
Ref.~\cite{BF}. Correspondingly, the
moment $\langle|q|\rangle$ of the overlap distribution~\cite{MPV} becomes
non-zero only at $p^*$, while the ground state energy and entropy remain
smooth for all $0\leq p\leq1$, even at the transition. Both, entropy
and overlap, exhibit an extremum at some $p>p^*$ which can be
explained in terms of the peculiar lattice hierarchy.

This study suggests that the reduction method provides a powerful
means to explore the (potential) onset of replica symmetry breaking
(RSB)~\cite{MPV} on finite-dimensional spin glasses for arbitrary bond
distributions at $T=0$ but variable $p$. The onset of glassy behavior
(albeit {\it not} RSB) in the Migdal-Kadanoff model at $p^*$, close to
and intrinsically linked with $p_c$, may suggest such a close link
between percolation and the possible onset of RSB in realistic lattice
models, which we are currently exploring \cite{BoPe4}. In fact, we
already found a transition for the stiffness exponent in real three-
and four-dimensional Edwards-Anderson models, resembling the one
observed here, at a $p^*>p_c$~\cite{eo_y}. 

Considered in terms of the average connectivity $\alpha=2dp\sim O(1)$
of spins, dilute lattice glasses as well as the hierarchical lattice
at $T=0$ behave strikingly similar to many mean-field models, which
are currently studied in the context of combinatorial
optimization~\cite{Monasson,Leone}. Those models also enter into a
glassy state (exhibiting RSB) at some finite connectivity $\alpha^*$,
above a finite-connectivity percolation transition
$\alpha_c<\alpha^*$. But in mean-field models the onset of RSB is
often studied as a precursor to yet another critical point, the
satisfiability transition \cite{AI}, whereas geometrically defined
models are unsatisfiable at any finite connectivity. These features
should be independent of the bond distribution, assuming zero mean and
unit variance.

While the hierarchical lattice is completely reducible, diluted
realistic lattices will become irreducible soon above percolation, in
which case our algorithm~\cite{eo_y} is used to reduce any graph as
much as possible first, followed by a complete exploration of the
ground states of the far more compact remainder graph using the
extremal optimization heuristic \cite{eo_prl}. With this hybrid
approach, we are typically able to compute ground state energies,
entropies and overlaps for graphs 10-100 times larger than previously
recorded, extending well beyond the percolation point.

\section{Migdal-Kadanoff Hierarchical Lattice}
\label{MKlat} 
To illustrate the reduction method, and to test our
algorithm, we consider here the hierarchical lattice (see
Fig.~\ref{hierlat}), obtained in the Migdal-Kadanoff real-space
renormalization scheme \cite{MK} for low-dimensional spin
glasses. These lattices have a simple recursive, yet geometric,
structure and are well-studied \cite{Kirkpatrick,SY,BF,BM,DM,BKM}. Most
importantly, they are completely reducible, and we can discuss the
method independently of any subsequent optimization that may be required
for more complicated models. The most interesting property of these
lattices is the curious fact that the scaling of its defect energy
distribution behaves very similar to that measured for actual two- and
three-dimensional lattices \cite{BM,F+H,eo_y}.

As described in Fig.~\ref{hierlat}, starting from generation $I=0$
with a single link, at each subsequent generation $I+1$, all links
from $I$ are replace with a new subgraph. The structure of the
subgraph arises from the bond-moving scheme in $d$ dimensions, and has
$2^d=8$ links for $d=3$ here. Thus, a hierarchical lattice of
generation $I$ has $l_I=(2^d)^I$ links, thus corresponding to a
$d$-dimensional lattice of ``length'' $L=2^I$ but
$n_I=2+2^{d-1}(L^d-1)/(2^d-1)=O(L^d)$ vertices. While the average
connectivity is $2l_I/n_I\sim4-2^{2-d}$, the two root-vertices from
generation $I=0$ themselves obtain in generation $I\gg1$ a
connectivity of $\sim2^{(d-1)I}$, and $\sim2^{dI-1}$ vertices are only
two-connected, i.~e. 7 in 8 for $d=3$.

The diluted hierarchical lattice percolates when there is a path
between the two root-vertices. This notion leads to a simple recursion
relation for the percolation threshold by counting the weights of all
diluted subgraphs from Fig.~\ref{hierlat} that percolate:
\begin{eqnarray}
p_{I+1}=4p_I^2-6p_I^4+4p_I^6-p_I^8,
\label{perceq}
\end{eqnarray}
which has a non-trivial stationary point at $p_c=0.2818376366$. It has
been pointed out by Ref.~\cite{BF} that a spin glass with the discrete
$\pm J$-bond distribution in Eq.~(\ref{bondeq}) exhibits instead a
critical transition between a paramagnetic and a spin glass phase at
$p^*=0.31032$, which is closely related to the percolation
transition. While below $p_c$ disconnected bonds ($J=0$) clearly
dominate and prevent long-range correlations, even for $p_c<p<p^*$
such correlations remain suppressed due to the cooperative behavior of
parallel bond structure pervasive in the lattice that lead to many
cancellations (see ``double rule'' in Sec.~\ref{rules}) and
additionally disconnects subgraphs at some higher level of the
hierarchy.

\section{Reduction Rules for Spin Glasses}
\label{rules}
In the Ising spin glass problem we assign to each vertex $i$ of a
graph or lattice, $1\leq i\leq n$, a spin variable $x_i\in\{-1,+1\}$,
while each link between two connected vertices $i$ and $j$ obtains a
bond variable $J_{i,j}$ that is drawn at random from some distribution
$P(J)$, which may be discrete or continuous. The problem consists of
finding spin configurations $(x_1,\ldots,x_n)$ which minimize the
Hamiltonian, or total energy,
\begin{eqnarray}
H(x_1\ldots,x_n)=-\sum_{<i,j>}\,J_{i,j}\,x_i\,x_j
\label{hamiltonianeq}
\end{eqnarray}
for a fixed (quenched) set of bonds $J_{i,j}$. Of course, for
$J_{i,j}>0$ (only ferromagnetic bonds), the solution is simply to make
all connected spins aligned with each other. But if some or all bonds
are negative, and spins are sufficiently connected, the problem can be
frustrated \cite{Toulouse} and solutions become nontrivial.

In the following, we explain the rules by which to eliminate
recursively all two-connected spins while accounting for the energy, entropy, and approximate overlap of ground-state
configurations. To this end we have to generalize bonds by adding
internal degrees of freedom, which evolve during the reduction
process. Each bond between two spins $i$ and $j$ now consists of a
tuple
\begin{eqnarray}
{\cal
J}_{i,j}=\left(J_{i,j},m_{i,j}^+,m_{i,j}^-,s_{i,j}^+,s_{i,j}^-\right),
\end{eqnarray}
where $J_{i,j}$ is the weight of the bond, $m_{i,j}^{\pm}$ is the
entropy accumulated by the bond, and $s_{i,j}^{\pm}$ counts the number
of previously reduced spins whose state is {\it completely
determined} (``entrained'') by the bond for $x_ix_j=\pm1$,
respectively. In addition, there is an energy offset $E_{\rm
offset}\leq0$ accounting for the energy difference between the
original and the reduced graph. In general, determining entropy and
overlap from the $m$- and $s$-values can be quite complicated for the
interesting case of an irreducible remainder graph~\cite{BoPe4}. For
the completely reducible hierarchical lattice, it is a lot simpler and
we will describe it below. Initially, for the unreduced graph, all
$m$- and $s$-values and $E_{\rm offset}$ are zero, and the bonds are
drawn from one of the usual distributions, for example, Eq.~(\ref{bondeq}).

These rules are elaborate but easily enumerated, {\it independent} of
the previous or future structure of the graph or the spin
configuration. A more complete set of rules including zero-, one-, and
three-connected vertices required for arbitrary graphs will be
presented elsewhere \cite{eo_y,BoPe4}. There, each spin $i$ obtains
one additional number $s_i$ (aside from its state $x_i$) that
counts the previously eliminated spins whose state is entrained to
$i$.

To demonstrate the use of these extra degrees of freedom, consider,
for instance, the reduction of a one-connected spin $j$: It and its
bond ${\cal J}_{i,j}$ to its sole neighbor $i$ disappears from the graph,
and the connectivity of $i$ is decremented. The new graph will have
itself a ground state energy offset by $|J_{i,j}|$ relative to the
unreduced graph, i.~e. $E_{\rm offset}\leftarrow E_{\rm
offset}-|J_{i,j}|$. For $J_{i,j}=0$ (a possibility we eventually have
to allow for, even if our unreduced bond distribution did not), all
information contained in ${\cal J}_{i,j}$ is discarded and $s_i$
doesn't change, since $x_j$ is free to take on any value $\pm1$ in the
ground state; only the entropy of the old graph is offset by $\ln2$
with respect to the new graph. In turn, for $J_{i,j}>0$ ($J_{i,j}<0$),
$j$ is entrained to the state of $i$ in any ground state
configuration, and $s_i$ is incremented. If the bond ${\cal J}_{i,j}$
had previously inherited nonzero entropies $m_{i,j}^{\pm}$ and
entrainments $s_{i,j}^{\pm}$, there would be an offset in entropy by
$m_{i,j}^+$ ($m_{i,j}^-$) between the reduced and the original graph,
and $s_i$ would be further increased by $s_{i,j}^+$ ($s_{i,j}^-$);
$m_{i,j}^-$ ($m_{i,j}^+$) and $s_{i,j}^-$ ($s_{i,j}^+$) are discarded
because the ground state {\it must} have $x_ix_j=+1$ ($x_ix_j=-1$).

With the two-point rule we use exclusively in this paper, a spin
variable $x_i$ and its two bonds ${\cal J}_1$ and ${\cal J}_2$ are
replaced by a new bond ${\cal J}$. It is easy to show that
\begin{eqnarray}
J=\frac{1}{2}\left(\left|J_1+J_2\right|-\left|J_1-J_2\right|\right),
\end{eqnarray}
which assigns to $J_{j,k}$ the bond with the smaller
modulus and the sign of $J_1J_2$. The energy offset $E_{\rm offset}$, accounting for the energy
of the original graph, will have to be lowered by
\begin{eqnarray}
\frac{1}{2}\left(\left|J_1+J_2\right|+\left|J_1-J_2\right|\right).
\label{h0eq}
\end{eqnarray}
At this level, our reduction rules correspond to the familiar traces
(at $T=0$) used many times before to study the hierarchical lattice
(see, e.~g., Refs.~\cite{DM,F+H}).
 
Clearly, for $|J_1|\not=|J_2|$, the reduced spin $i$ is entrained to the
neighbor with the bond of larger modulus, while the weaker bond may be
violated. The entropy and entrainment variables that ${\cal J}$
inherits from the two reduced bonds are controlled by the sign of the
stronger one. We can summarize these rules by defining ${\cal J}_>$
and ${\cal J}_<$ as the bond with the stronger and weaker (in modulus)
weight, respectively. Then, we obtain for the new bond:
\begin{eqnarray}
\underline{|J_>|>|J_<|}&:&\nonumber\\
m^{\pm}&=&m_>^{sign(J_>)}+m_<^{\pm sign(J_>)},\nonumber\\
s^{\pm}&=&s_>^{sign(J_>)}+s_<^{\pm sign(J_>)}.
\label{2rulenoteq}
\end{eqnarray}
The case $|J_1|=|J_2|$ leads to further distinctions:
\begin{eqnarray}
m^{\pm}&=&
\begin{cases}
m_1^{sign(J_1)}+m_2^{sign(J_2)}&,J_1=\pm J_2\not=0\\
\ln\left(e^{m_1^++m_2^{\pm}}+e^{m_1^-+m_2^{\mp}}\right)&,J_1=\mp J_2
\end{cases}
\nonumber\\ 
s^{\pm}&=&
\begin{cases}
s_1^{sign(J_1)}+s_2^{sign(J_2)}&,J_1=\pm J_2\not=0\\
0&,J_1=\mp J_2
\end{cases}
\label{2ruleequaleq}
\end{eqnarray}

Finally, we need a rule to combine double bonds, which will be created
eventually. When two bonds, ${\cal J}_1$ and ${\cal J}_2$, connect
the same two spins, they are simply replaced with a new bond
\begin{eqnarray}
{\cal J}={\cal J}_1+{\cal J}_2.
\label{doubleruleeq}
\end{eqnarray}
It is important to note that the two-point and double rules only
effect the newly created bond and the energy offset, but not the two
neighboring spins (or the entropy offset). Unlike for the one-point
rule, we do not need to consider an internal entrainment number for
spins here.

\section{Algorithm and Implementation}
\label{algo}
With the reduction rules from Sec.~\ref{rules}, it is now
straightforward to evaluate hierarchical lattices of any size.  In
effect, these rules allow us to reverse the generating mechanism
depicted in Fig.~\ref{hierlat}. Recursively, from one generation to
the next, we reduce all two-connected spins in the middle of the
subgraph first, then use the double-bond rule sequentially, until a
single, reduced bond is left. Once an entire graph has been reduced
into such a single bond, the weight $J$ of that bond is in fact half
the defect energy, $\Delta E=2J$, of that graph, i.~e. the energy
difference between having both root spins aligned and
anti-aligned. The entropy of that graph is simply given by
$m^{sign(J)}$, or by $\ln[\exp(m^+)+\exp(m^-)]$ for $J=0$. The energy
of the graph can be obtained from the running energy-offset $E_{\rm
offset}$, accounting for the reduction of bonds along the way, and
$J$. For our purposes, it is more useful to extract the (non-negative)
``cost'' of a graph given by the absolute weight of all violated bonds
for any bond distribution. In case of the discrete $\pm J$
distribution in Eq.~(\ref{bondeq}), this sum reduces to a count of all
violate bonds (all of absolute weight 1) $(l+E_{\rm offset}-|J|)/2$,
where $\langle l\rangle=pL^d$ is the average number of non-vanishing
bonds in a graph. As an approximate measure of the overlap, we merely
store the largest entrainment number $s_{\rm max}$ observed at any
point during the reduction process, which is a measure of the largest
correlated cluster of spins within the lattice. The size of this
cluster will either dominate the overlap (with smaller clusters adding
some negligible fluctuations) in the spin glass state or be itself a
vanishing fluctuation in the paramagnetic state.

Although these measurements are not difficult, the reduction of ever
larger graphs soon becomes untenable, since the cost of reducing true
graphs would grow by a factor of $2^d$ for each generation. Instead,
we exploit the fact that each subgraph within the hierarchy is
independent of any other parallel subgraph. As above, we reduce all
graphs down to a single bond. To assemble a graph of generation $I+1$
we only need a large enough pool of independent graphs of generation
$I$. In fact, {\it many nearly-independent} graphs of generation $I+1$
can be assembled by repeatedly drawing at random on $2^d$ graphs of
generation $I$ from that pool.

We proceed as follows: Assume that we already possess a pool of $A_I$
reduced graphs at every generation $I$ (up to a maximal generation
$I_{\rm max}$), each graph represented by its bond ${\cal J}$, $E_{\rm
offset}$, and $s_{\rm max}$. Each cycle of the algorithm generates an
elementary subgraph (generation $I=1$) with bonds $J$ drawn from the
bond distribution in Eq.~(\ref{bondeq}) and empty entropy,
entrainment, $E_{\rm offset}$, and $s_{\rm max}$ variables. The
elementary subgraph is reduced to obtain a bond ${\cal J}$ of the
first generation, {\it replacing} the oldest entry in pool
$A_1$. After every $k$th new reduced bond entering into pool
$A_{I-1}$, we randomly assemble $2^d$ such bonds into a new subgraph
to create a new reduced bond to be entered into pool $A_I$, as long as
$I\leq I_{\rm max}$.

If we choose $A_I=k=2^d$ and assemble bonds sequentially instead of at
random, this algorithm produces exact hierarchical lattices, where
each subgraph is certain to be independent. But to obtain just one
graph of, say, generation $I=10$, we have to assemble $2^{30}$
elementary subgraphs for $d=3$. Instead, for all $p$ not too close to
$p^*$, we simply choose large pools, $A_I=2048$, and as small as
$k=1$. That is, each time we assemble an elementary graphs at $I=0$,
we enter its value to pool $A_1$, assemble a subgraph out of $A_1$,
add the reduced bond to pool $A_2$, etc, all the way up to $I_{\rm
max}$ in each cycle. Hence, our implementation becomes {\it
independent} of $L$, and for instance at $p=1$ we have assembled
graphs up to generation $I_{\rm max}=100$ in $d=3$ and $I_{\rm
max}=35$ in $d=4$, corresponding to $L=2^{100}$ or $2^35$,
respectively.  $I_{\rm max}$ is only limited by apparent limits on
floating point accuracy due to repeated cancellation between almost
equal bonds. Note that each pool gets completely refreshed after $A_I$
creations of new graphs from the pool just below it, while there are
about $O(2^{dA_I})$ different subgraphs that could be created out of
each pool at any one time.

Every time a graph gets added to a pool, its properties are evaluated
from its ${\cal J}$, $E_{\rm offset}$, and $s_{\rm max}$. While ground
state energy, entropy, and (approximate) overlap converge quickly with
$I$, and throughout are quite independent of the choice of $k$ and
$A_I$, the measurement of the defect energy becomes quite sensitive to
the algorithmic parameters near $p^*$. In those cases, we increase $k$
(and lower $I_{\rm max}$ correspondingly), until stable results were
obtained. Only at $p\approx p^*$, the exact algorithm was necessary to
observe the true defect energy distribution.

\begin{figure}
\vskip 2.2in \includegraphics{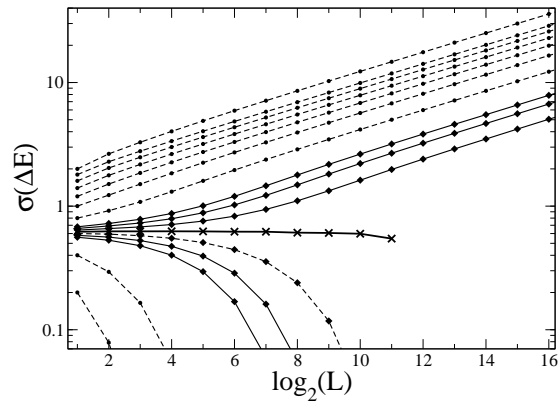}
\caption{Plot of the width $\sigma(\Delta E)$ of the defect energy
distribution as a function of systems size $L$ in $d=3$ for various
bond fractions $p$. Dashed lines from bottom to top refer to
$p=0.1,0.2,\ldots,1$, while data with diamonds and crosses refer to
$p=0.28,0.29,\ldots,0.34$. The data points with circles were obtained
using the $k=1$ implementation from Sec.~\protect\ref{algo}, those
with diamonds using $k=2$, and those with crosses at $p=0.31\approx
p^*$ required the exact algorithm. Note that for $p<p^*$ the width
evolves toward the $p=0$ fix-point with $\sigma(\Delta E)=0$, while
for $p>p^*$ it invariably evolves to the $p=1$ fix-point with scaling
behavior $\sigma(\Delta E)\sim L^y$. While the exact asymptotic
behavior emerges only for very large sizes $L$ when $p\approx p^*$, it
is already obvious for small $L$ whether $\sigma(\Delta E)$ is going
to rise or to fall.}
\label{defectscal}
\end{figure}

\section{Numerical Results}
\label{results}
Using the algorithm from Sec.~\ref{algo}, we have studied various
properties of a Ising spin glass with bond distribution as given in
Eq.~(\ref{bondeq}) on the diluted and undiluted hierarchical
lattice. In fact, using the easiest implementation with $k=1$, pool
sizes of $A_I=2048$, and $I_{\rm max}=20$, we first scanned lattices
for $0<p\leq1$ in steps of $\Delta p=0.1$. At each $p$ and $I$, we
averaged over $10^7$ values. We found that for each $p$, the ground
state energies per spin, entropies per spin, and the moment of the
overlap, $\langle|q|\rangle$ were well-converged at $I_{\rm max}=20$,
and insensitive to the choice of $k$ using trials with $k=2$.

We also found the width of the defect energy distribution,
$\sigma(\Delta E)=\sqrt{\langle\Delta E^2\rangle-\langle\Delta
E\rangle^2}$, to be stable except for $p=0.3$. For large $I=\log_2(L)$
the distribution of defect energies either collapses into a
$\delta$-function at zero for small $p$, or eventually converges to
the distribution of the undiluted lattice ($p=1$). Near $p\approx
p^*$, the evolution of that distribution becomes highly sensitive to
small fluctuations, and the minute statistical dependencies inherent
in our algorithm for small $k$ escalate. Thus, we ran another scan for
$0.28\leq p\leq0.34$ using $k=2$ and $10^4$ graphs at $I_{\rm max}=20$
(and hence gaining a factor $k$ more graphs at each lower generation
$I$). Now, only for $p=0.31\approx p^*$ did $\sigma(\Delta E)$ become
unstable (by comparison with trial runs at $k=4$). Finally, for
$p=0.31$, we used an exact algorithm and indeed found $\sigma(\Delta
E)$ to be horizontal for many generations until it eventually decays,
indicating that $p^*\gtrsim0.31$, consistent with the expected value
of $0.31032$ from Ref.~\cite{BF}. The results for $\sigma(\Delta E)$
as a function of $I=\log_2(L)$ for all $p$ are shown in
Fig.~\ref{defectscal}.

\begin{figure}
\vskip 2.0in \includegraphics{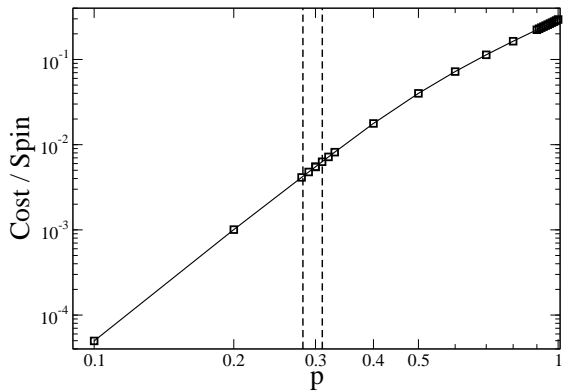}
\caption{Log-log Plot of the cost per spin for the infinite system in
$d=3$ as a function of $p$. This cost, i.~e. the fraction of violated
bonds, only approaches zero for $p\to0$ (consistent with $p^4$). The
data smoothly continues through both critical points at $p_c$ and
$p^*$, indicated by dashed vertical lines. For $p>p^*$ the cost
appears to rise less rapidly than for $p<p^*$. }
\label{mkenergy}
\end{figure}

\begin{figure}
\vskip 2.0in \includegraphics{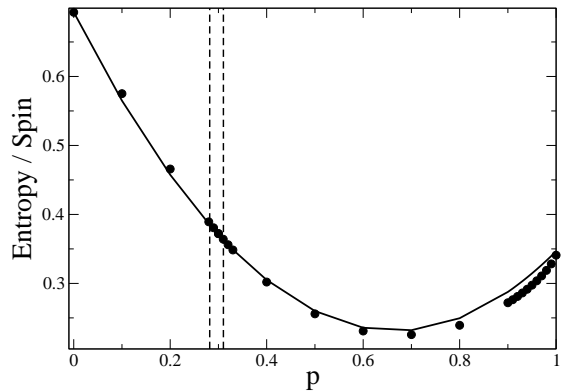}
\caption{Plot of the entropy per spin for the infinite system as a
function of $p$. The data smoothly continues through both critical
points at $p_c$ and $p^*$, indicated by dashed vertical lines, but
exhibits a surprising minimum. The data is approximated well by
Eq.~(\protect\ref{Seq}) (continuous line), as discussed in the text.}
\label{mkentropy}
\end{figure}

\begin{figure}
\vskip 2.0in \includegraphics{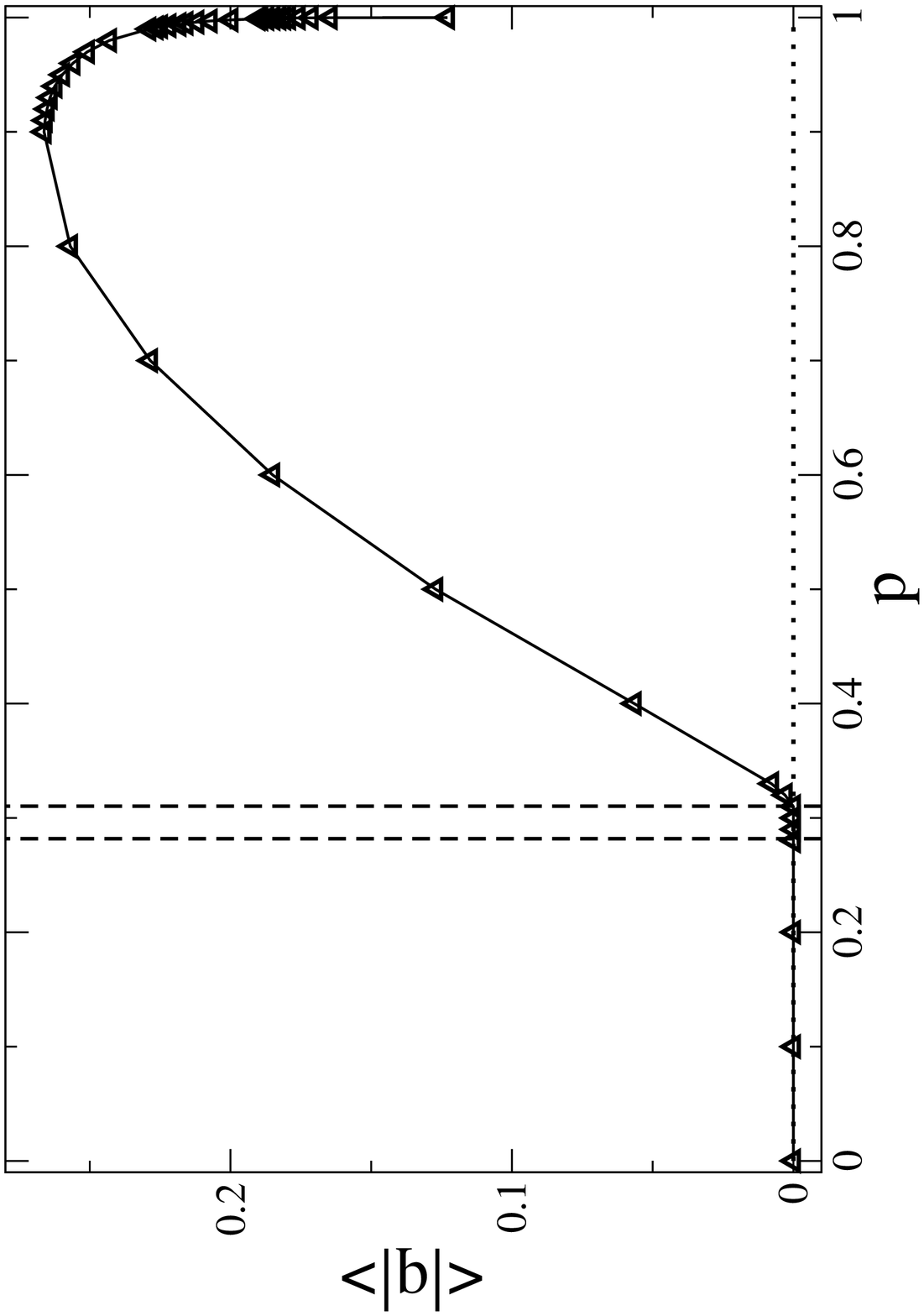} \includegraphics{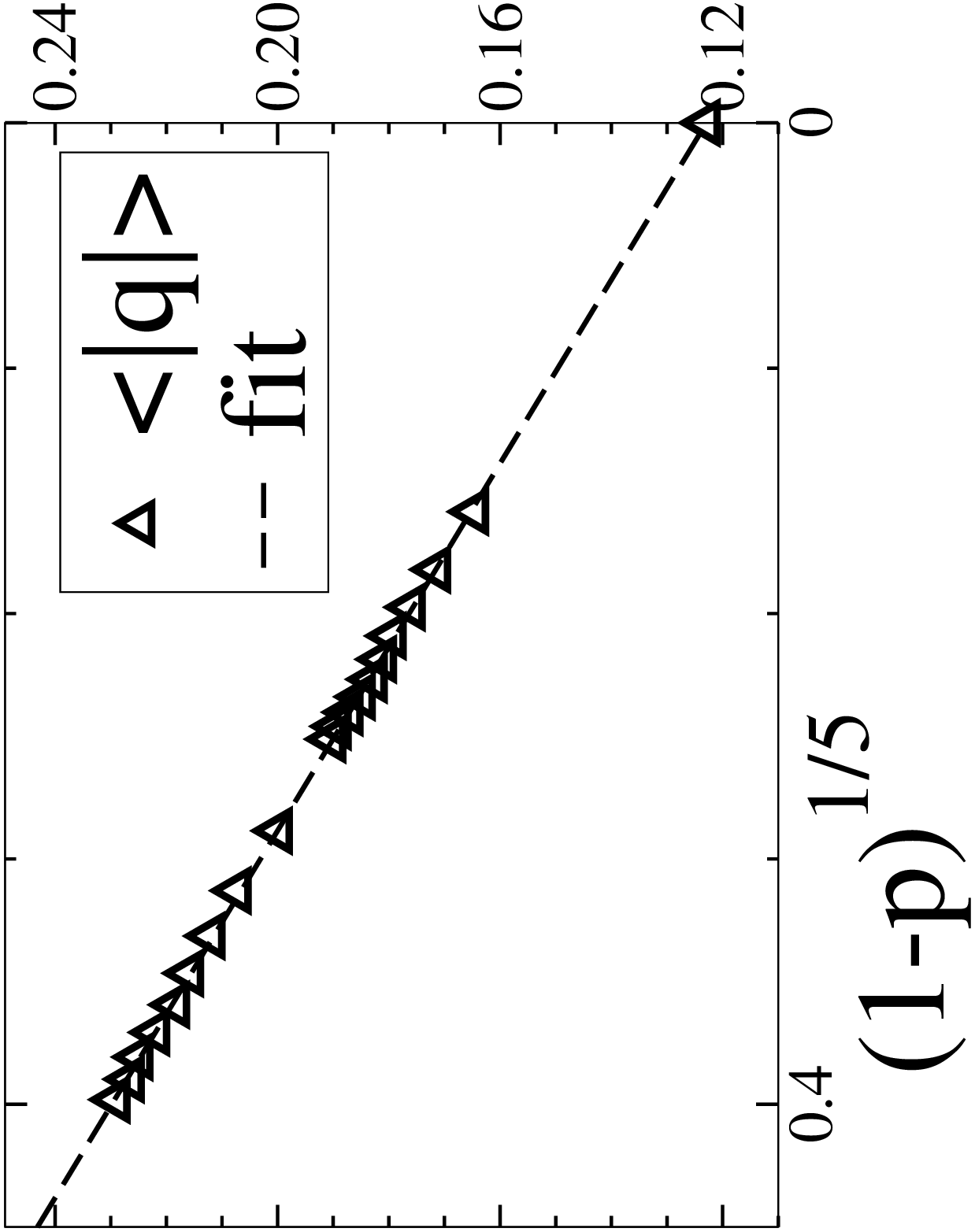}
\caption{Plot of the moment $\langle|q|\rangle$ of the overlap
distribution as a function of $p$. The moment remains zero for all
$p\leq p^*$, including at the percolation point $p_c$. (Both, $p_c$
and $p^*$, are marked by dashed vertical lines.) Above the transition
at $p^*$ to a spin glass state, $\langle|q|\rangle$ rises first
linearly, but then turns over at a maximum at small dilutions just
below $p=1$, and drops sharply at $p=1$. The inset shows that
$\langle|q|\rangle$ seems to scale with $(1-p)^{1/5}$ for $p\to1$; the
dashed line is a linear fit. }
\label{mkoverlap}
\end{figure}

In Figs.~\ref{mkenergy}-\ref{mkoverlap} we plot the number of violated
bonds (or cost) per spin, entropy per spin, and a characteristic
moment of the (approximate) overlap distribution for $L\to\infty$ and
$d=3$. The cost is already non-zero for small $p$, rising consistent
with $p^4$, proportional to the probability of creating a small,
frustrated loop that requires four bonds. Just beyond $p^*$, the cost
per spin rises more slowly, but without drastic changes for any
$p$. Similarly, the entropy density in Fig.~\ref{mkentropy} remains
unaffected by the transition at $p^*$, as is expected
\cite{Zecchina}. Instead, it reaches a minimum at about $p=0.7$ that
is an artifact of the hierarchical lattice. The large majority of
spins in the graph (a fraction $\sim 7/8$ for $d=3$) are at most
two-connected spins whose contribution dominates the entropy
density. In a fully bonded graph ($p=1$), about half of these spins
are free, the other half is constraint, with an energy density of
$\ln(2)/2$. For small dilutions $p\lesssim1$, about every removed bond
turns a two-connected spin into a one-connected spin, which is {\it
always} constrained, and the entropy declines. Eventually, at large
dilution (small $p$), an increasing number of one-connected spins
become totally disconnected and free again, raising to entropy density
to its ultimate value, $\ln(2)$, for the unconnected ($p=0$) lattice.
By this argument the entropy density should be given simply by
\begin{eqnarray}
\frac{S}{n}\approx\left[(1-p)^2+\frac{1}{2}\,p^2\right]\ln(2),
\label{Seq}
\end{eqnarray}
which fits the data exceedingly well.

In contrast with energy and entropy, the overlap distribution is
clearly sensitive to the transition from paramagnetic to glassy
behavior at $p^*$, as shown in Fig.~\ref{mkoverlap}. There, we plot
only its moment, since the distribution itself quickly converges to a
structureless $\delta$-peak as expected for a model that is replica
symmetric (RS) throughout. In fact, we find that
$\langle|q|\rangle/\sqrt{\langle q^2\rangle}\sim1$ to within 1\% for
all $p$. (For one, a broad overlap distribution indicative of widely
separated ground state configurations and RSB~\cite{MPV} would be
antithetical to the reducibility of the hierarchical lattice.) About
at $p^*$, the location of the peak becomes non-zero and splits into
two symmetric peaks, leading to a non-zero moment $\langle|q|\rangle$.

The moment experiences a surprising maximum very close to $p=1$,
varying sharply for small dilutions. Again, the maximum may be
explained in terms of the abundance of two-connected spins: Whatever
the size of strongly correlated spin clusters may be in the fully
connected lattice ($p=1$), the overwhelming conversion of
two-connected to one-connected spins at small dilution can only
increase the size of those clusters at its edge, since one-connected
spins are always entrained. But more importantly, the elimination of
two-connected spins, say, within a subgraph as shown in
Fig.~\ref{hierlat} will {\it enhance} the probability that the two
root-spins of that subgraph become correlated, which would {\it merge}
the clusters that those spins are part of. While the latter effect
could suddenly increase the size of typical clusters by a factor, it
is not clear whether the steep slope observed at $p=1$ is not an
artifact of our approximation to the overlap (see
Sec.~\ref{algo}). Further numerical studies suggest that
\begin{eqnarray}
\langle|q|\rangle_p-\langle|q|\rangle_{p=1}\sim(1-p)^{\frac{1}{5}}\quad(p\to1),
\label{qeq}
\end{eqnarray}
as seen in the inset of Fig.~\ref{mkoverlap}. We have not found an
explanation for the scaling in Eq.~(\ref{qeq}) in terms of the lattice
yet, since it is also a property of the entrainment in the spin glass
state (and hopefully, the overlap).

As shown in Fig.~\ref{defectscal}, for all $p>p^*$ we obtain the
expected scaling of the width of the defect energy distribution with
system size $L$,
\begin{eqnarray}
\sigma(\Delta E)\sim a\,L^y\,\left[1+\epsilon(L)\right],
\label{scaleq}
\end{eqnarray}
for large enough $L$, including a next-to-leading correction
$\epsilon(L)\ll1$. Most conveniently, it appears that our $k=1$
algorithm (for large enough pools $A_I$) described in Sec.~\ref{algo}
is least likely to create spurious dependencies for $p=1$, where we
would expect the quickest and most stable convergence toward the
renormalized bond distribution. Thus, we disabled the sampling of
energy-offsets, entropies, and entrainments, which already are
well-converged at $I_{\rm max}=20$, and only focused on the reduction
of bonds. Setting $I_{\rm max}=100$ ($d=3$) or $I_{\rm max}=40$
($d=4$) , $k=1$, and $A_I=1024$ for each $I$, we measured more than
$10^9$ reduced bonds {\it at each} $I$ to obtain the distributions of
defect energies $\Delta E=2J$ in $d=3$ and~4. (In fact, we repeated
the same calculation with a different random number generator, without
markable difference in results. A previous run with $I_{\rm max}=150$
in $d=3$ showed divergence due to limited floating point precision at
$I\approx120$; the same problem emerged already at $I_{\rm max}=40$ in
$d=4$.)  At that level, we obtained a ratio between its absolute mean
and its width $\sigma(\Delta E)$ of $<10^{-4}$ in both dimensions,
even for the largest $I$, which is a measure of the statistical error
(the distribution is supposed to be symmetric).

\begin{figure}
\vskip 4.2in 
\includegraphics{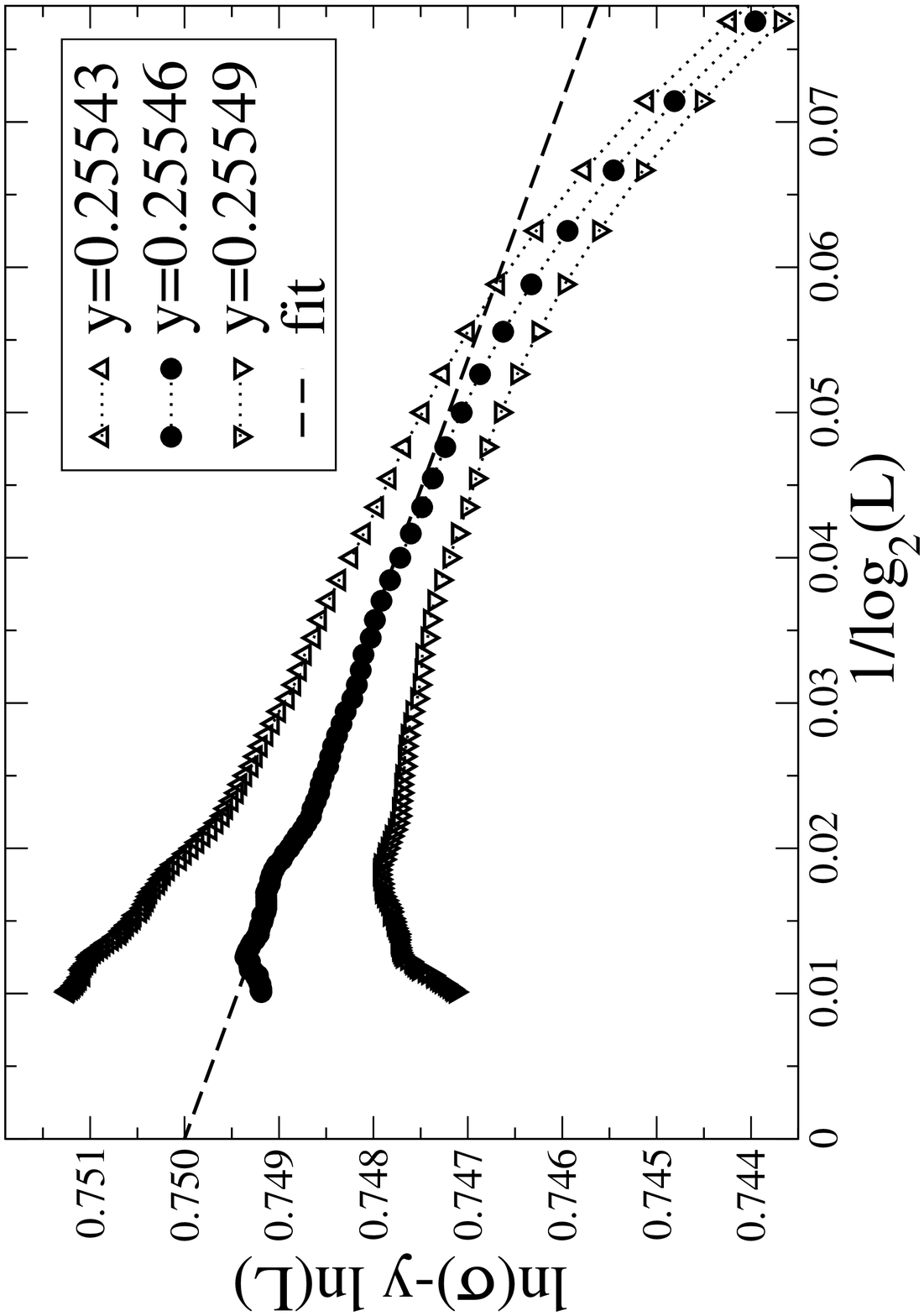}
\includegraphics{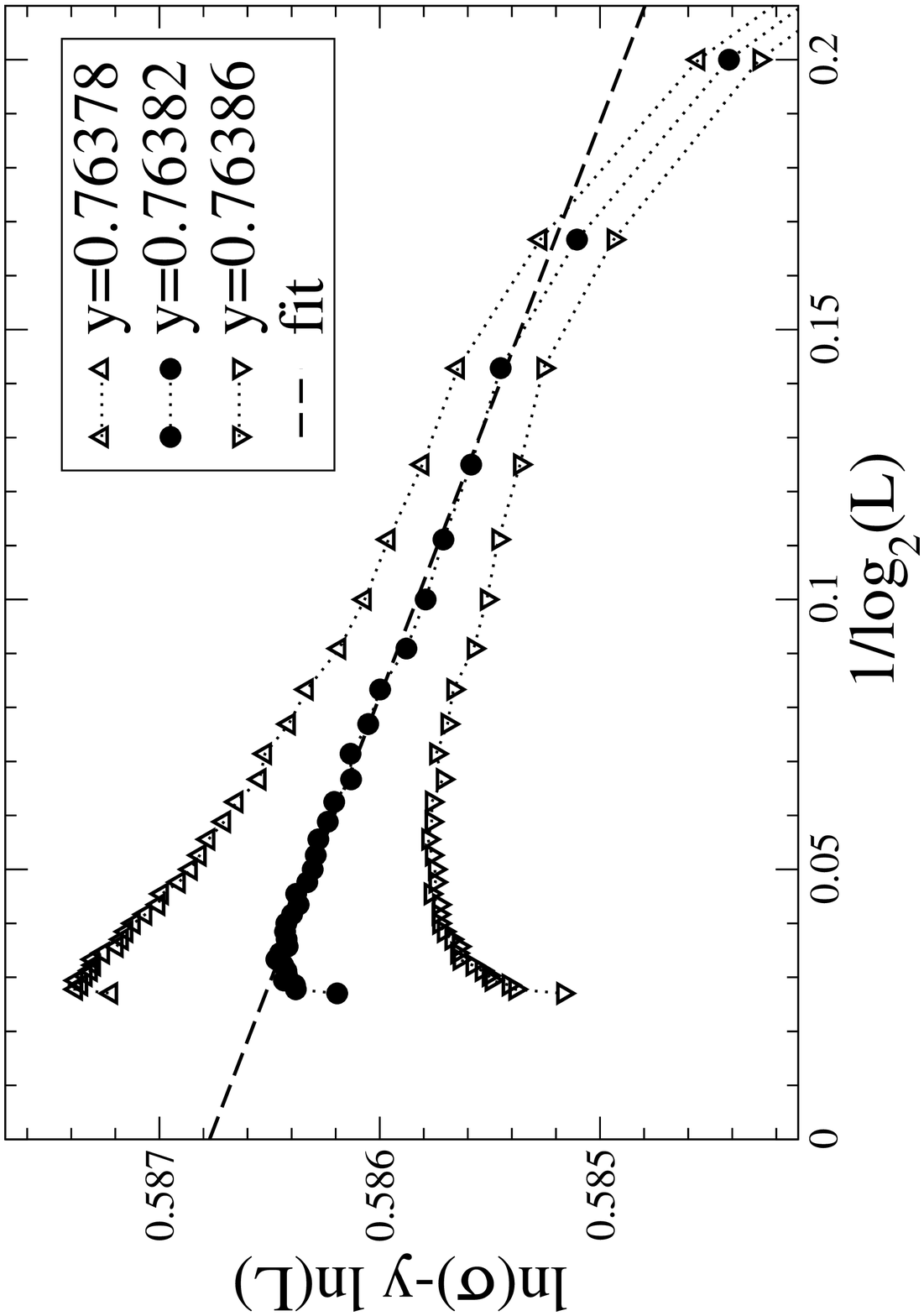}
\caption{Extrapolation plot for the leading asymptotic behavior of
$\sigma(\Delta E)$ as a function of system size $L$ in $d=3$ (top) and
$d=4$ (bottom). The data for $\sigma(\Delta E)$ is converted to
$\ln(\sigma)-y\ln(L)$ and plotted vs. the apparent scaling correction,
$1/\log_2(L)$, for different estimates of the exponent $y$. For $d=3$
the graph at the center (circles) plots the data for $y=0.25546$ and
seems to converge linearly to a finite value of about $3/4$ for
$L\to\infty$. In turn, using $y=0.25549$ (down triangle) or
$y=0.25543$ (up triangle) on the same data seems to lead tor diverging
extrapolations, suggesting $y_3=0.25546(3)$. The same procedure for
$d=4$ brackets the ``best'' choice of $y=0.76382$ (circles) between
two diverging choices of $y=0.76386$ (down triangle) and $y=0.76378$
(up triangle), suggesting $y_4=0.76382(4)$. }
\label{yextra}
\end{figure}

Surprisingly, although we extended the scaling regime in $d=3$ by a
{\it power} of 10 compared to previous calculations ($L=2^{100}$
compared to $L=2^{10}$ \cite{DM}), the extrapolation for the scaling
exponent $y$ in Eq.~(\ref{scaleq}) only yields 5 digits accuracy, see
top of Fig.~\ref{yextra}. First, we approximated $y\approx0.2555$ from
the intercept of $\log_2(\sigma)/\log_2(L)$ with respect to
$1/\log_2(L)$ at $L\to\infty$, which displays large corrections to
scaling. We are then able to refine our estimate of $y$ by converting
Eq.~(\ref{scaleq}) to $\ln(\sigma)-y\ln(L)\sim\ln(a)+\epsilon(L)$,
using $\ln(1+\epsilon)\sim\epsilon$. Plotting that expression again
vs. $1/\log_2(L)$ (note that $\log_2(L)=I$) suggest a straight-line
asymptotic behavior that would imply $\epsilon\sim b/\log_2(L)$ for
$L\to\infty$. The top of Fig.~\ref{yextra} indicates that a value of
$y=0.25546(3)$ in $d=3$ leads to an apparently converging result with
$\ln(a)\approx3/4$ at the intercept for $L\to\infty$. We have used the
same procedure to find $y=0.76382(4)$ in $d=4$, see bottom of
Fig.~\ref{yextra}. Although the floating point accuracy here
deteriorated already at $L=2^{35}$, the asymptotic scaling regime is
reached for much small $L$ than in $d=3$, leading to comparable
accuracy in the fit.

Our value for the stiffness exponent $y$ in $d=3$ is 6\% below
$y=[d-1+\log_2(1-2/\pi)]/2=0.2698$ obtained analytically in
Ref.~\cite{SY} with a Gaussian approximation, and consistent with
$y\approx0.25$ obtained numerically in Ref.~\cite{BM} by evolving a
``pool'' of bonds through several generations, very similar to our
algorithm in Sec.~\ref{algo}. Since both of these references use an
initial Gaussian bond distribution, in contrast to our discrete
distribution in Eq.~(\ref{bondeq}), the results suggest that the
stiffness exponent is independent of the bond distribution. In
Ref.~\cite{BKM}, an expansion in the limit of large dimension
$d\to\infty$ is used to demonstrate the independence of $y$ of the
initial bond distribution. A recursion in the generation $I$ is
constructed which for $I\to\infty$ leads invariably to the Gaussian
derived in Ref.~\cite{SY}. With finite-dimensional corrections, this
expansion improves on the Gaussian result above to yield
$y_3\approx0.255$, in excellent agreement with our result here. In
contrast, we expect that energy, entropy, overlap, and even $p^*$
would depend on the distribution at least quantitatively.

\section{Conclusions}
\label{conclusion}
We have presented a new method to reduce spin glasses on arbitrary
graphical structures. The reductions are most appropriate for sparse
graphs near their percolation threshold, which may be either
completely (as in this paper) or substantially reduced, facilitating
further optimization using standard methods \cite{eo_y,BoPe4}. In
either case, the reductions allow to extract exact ground state
energies and entropies, and even provide a good approximation to
collective properties such as the overlap distribution.

With the reduction method we have reproduced a number of ground-state
properties of the Migdal-Kadanoff hierarchical lattice at any
dilution, and discovered various new features, which can be mostly
explained in terms of the lattice structure. The simplicity of this
model further permitted us to determine the exponent $y$ in $d=3$
and~4 for the growth of the defect energy with system size to
unprecedented accuracy. We have recently applied the reduction method
to diluted Edwards-Anderson models for $d=3$ and~4, obtaining
$y_3=0.240(5)$ and $y_4=0.60(1)$ for the respective stiffness
exponents~\cite{eo_y}. These results demonstrate that the
Migdal-Kadanoff approximation in $d=3$ is even closer to the
Edwards-Anderson value than had been previously
expected~\cite{BM,F+H}, although they are definitely distinct. The
differing values in $d=4$ reflect the fact that Migdal-Kadanoff is a
low-dimensional approximation.

 The results obtained here and in Ref.~\cite{eo_y} suggest that the
reduction method may be a useful tool in the exploration of Ising spin
glass systems on any graph or lattice near percolation that may reveal
the onset of non-trivial configurations of ground states.

\section*{Acknowledgments}
I would like to thank A.~Percus and M.~Paczuski for helpful discussions.

\end{document}